\documentclass[amssymb,showkeys,12pt]{revtex4-1}
\usepackage{graphicx}
\usepackage{multirow}
\usepackage{float}
\usepackage{bm}
\usepackage{amsmath}
\usepackage{amssymb}
\usepackage{hyperref}
\begin{document} 
\title{\normalsize Tunneling time in attosecond experiments,  
Keldysh, Mandelstam-Tamm and intrinsic-type of time.
}
\author{Ossama Kullie} 
\affiliation{Theoretical Physics, Institute for Physics, Department 
of Mathematics and Natural Science, University of Kassel, Germany}
\thanks{\tiny Electronic mail: kullie@uni-kassel.de}
\begin{abstract} 
\scriptsize
Tunneling time in attosecond and strong field experiments is one of the most 
controversial issues in today's research, because of its importance to the theory 
of time, the time operator and the time-energy uncertainty relation in 
quantum mechanics.   
In \cite{Kullie:2015I} we derived an estimation of the (real) tunneling 
time, which shows an excellent agreement with the time measured in attosecond 
experiments, our derivation is found by utilizing the time-energy uncertainty 
relation, and it represents a quantum clock. 
In this work, we show different aspects of the tunneling time in attosecond 
experiments, 
we discuss and compare the different views and approaches, which are 
used to calculate the tunneling time, i.e. Keldysh time (as a real or imaginary 
quantity), Mandelstam-Tamm time and our tunneling time relation(s). We draw some 
conclusion concerning the validity and the relation between the different 
types of the tunneling time with the hope, it will help to answer the the 
question put forward by Orlando et al \cite{Orlando:2014I,*Orlando:2014II}: tunneling time, what does it mean?.
In respect to our result, the time in quantum mechanics can be, in more general 
fashion, classified in two types, intrinsic dynamically connected, and external  
dynamically not connected, to the system.     
\end{abstract} 
\maketitle     
\scriptsize
\subsection{Introduction}\label{ssec:int}                
The area of time-energy uncertainty relation (TEUR) continues to attract
the attention of many researchers until now, and it remains alive
almost 90 years after its birth \cite{Dodonov:2015}. It received, since 1980s, 
a ‘new breath’  due to the actual problems of quantum information theory and 
impressive progress of the experimental technique in quantum optics and atomic 
physics \cite{Dodonov:2015,Maquet:2014}.
There is no doubt that the advent of ‘attophysics’ opens new perspectives 
in the study of time resolved phenomena in atomic and molecular physics 
\cite{Maquet:2014}. 
Since the appearance of the quantum mechanic the time  was controversial, 
the famous example is the Bohr-Einstein weighing 
{\it photon box Gedanken experiment}. The crucial point is that, to date,  
there is no general time operator found, thus the uncertainty relation is used 
dependent upon the study case and usually reduced to the 
position-momentum uncertainty relation or used in a pragmatic way 
\cite{Muga:2008} (chap. 3).   
There is still  common opinion that time plays a role 
essentially different from the role of the position in quantum mechanics 
(although it is not in line with special relativity), and that the time is not 
an operator but a parameter and hence does not obey an ordinary time-energy 
uncertainty relation (TEUR).
Busch \cite{Busch:1990I,Busch:1990II} argued that the conundrum of the TEUR in 
quantum mechanics is related in first place to the fact  that the time is 
identified as a parameter in Schr\"odinger equation (SEQ).
Hilgevoord \cite{Hilgevoord:2002} has showed that there is nothing in the 
formalism of the quantum mechanics that forces us to treat time and position 
differently. Observables such as position, velocity, etc. both in classical 
mechanics as well as in quantum mechanics, are relative observables, and one 
never measures the absolute position of a particle, but the distance in between 
the particle and some other object \cite{Hilgevoord:2002,Aharonov:2000}. 
Hilgevoord concluded \cite{Hilgevoord:2002} that when looking to a time operator  
a distinction must be made between universal time coordinate $t$, a c-number like 
a space coordinate, and the dynamical time variable of a physical system suited 
in space-time, i.e. clocks.

Indeed there have been many attempts to construct time as an operator 
\cite{Razavy:1967,Goto:1981,Han:1983,Wang:2007}, Busch 
\cite{Busch:1990I,Busch:1990II} classified three types of time in quantum 
mechanics: external time (parametric or laboratory time), intrinsic or 
dynamical time and observable time. 
External time measurements are carried out with clocks that are not dynamically 
connected with the object studied in the experiment, and usually called 
parametric time. 
The intrinsic-type of time is measured in terms of the physical system 
undergoing a dynamical change, the time is defined through the dynamical 
behavior of the quantum object itself, where every  dynamical variable marks the 
passage of time.
The third type, according to Busch, is the observable time or event time, for 
example the time of arrival of the decay products at a detector. 
One of the most impressive debates about the time and the TEUR in quantum mechanic 
is the (Bohr-)Einstein weighing
\cite{Cooke:1980,Aharonov:2000}, or short, Bohr-Einstein {\it Gedanken experiment} 
({\it Bohr-Einstein-GE}).
A photon is allowed to escape from a box through a hole, which is closed and 
opened temporarily  by a shutter, the period of time is determined by a clock, 
which is part of the box system, which means that the time is intrinsic and 
dynamically connected with the system under consideration. 
 The total mass of the box before and after a photon passes is measured. 
Bohr showed that the process of weighing introduces a quantum uncertainty 
(in the gravitational field) which leads to an uncertainty in time $\tau$, 
the time needed to pass out of the box that usually called passage time 
\cite{Busch:1990I,Busch:1990II}, in accordance with the TEUR: 
\begin{equation}\label{ucr} 
  {\,\, \Delta T  \cdot \Delta E \ge \frac{\hbar}{2}}
\end{equation}
Aharonov and Rezinik \cite{Aharonov:2000} offer a similar interpretation, that
the weighing leads, due to the back reaction of the system underlying  
a perturbation (energy measurement), to an uncertainty in the time of the 
internal clock relative to  the external time \cite{Aharonov:2000}. 
Hence for quantum systems it is decisive to observe the time from within 
the system or using an internal clock. 
\vspace*{-0.50cm}
\subsection{Basic concepts of the attosecond experiment}\label{ssec:bc} 
In attosecond experiment the idea is to control the 
electronic motion by laser fields that are comparable in strength to the electric 
field in the atom. In today's experiment usual intensities are 
$\sim 10^{14}\, W cm^{-2}$, for more details we refer to the tutorial 
\cite{Krausz:2009,Dahlstrom:2012}. 
In  the majority of phenomena in attosecond physics, one can separate 
the dynamics into a domain ''inside'' the atom, where atomic forces dominate, 
and ``outside``, where the laser force dominates, a two-step semi-classical
model, pioneered by Corkum \cite{Corkum:1993}.
Ionization as the transition from ''inside'' to ''outside'' the atom plays a key 
role for attosecond phenomena. A key quantity here is the Keldysh parameter 
$\gamma_{_K} = \frac{\sqrt{2I_p}}{F} \omega$, where $I_p$ denotes the ionization 
potential of the system (atom or molecule)  
\begin{figure}[t]
\vspace{-4.5cm}
\includegraphics[height=16.0cm,width=13cm]{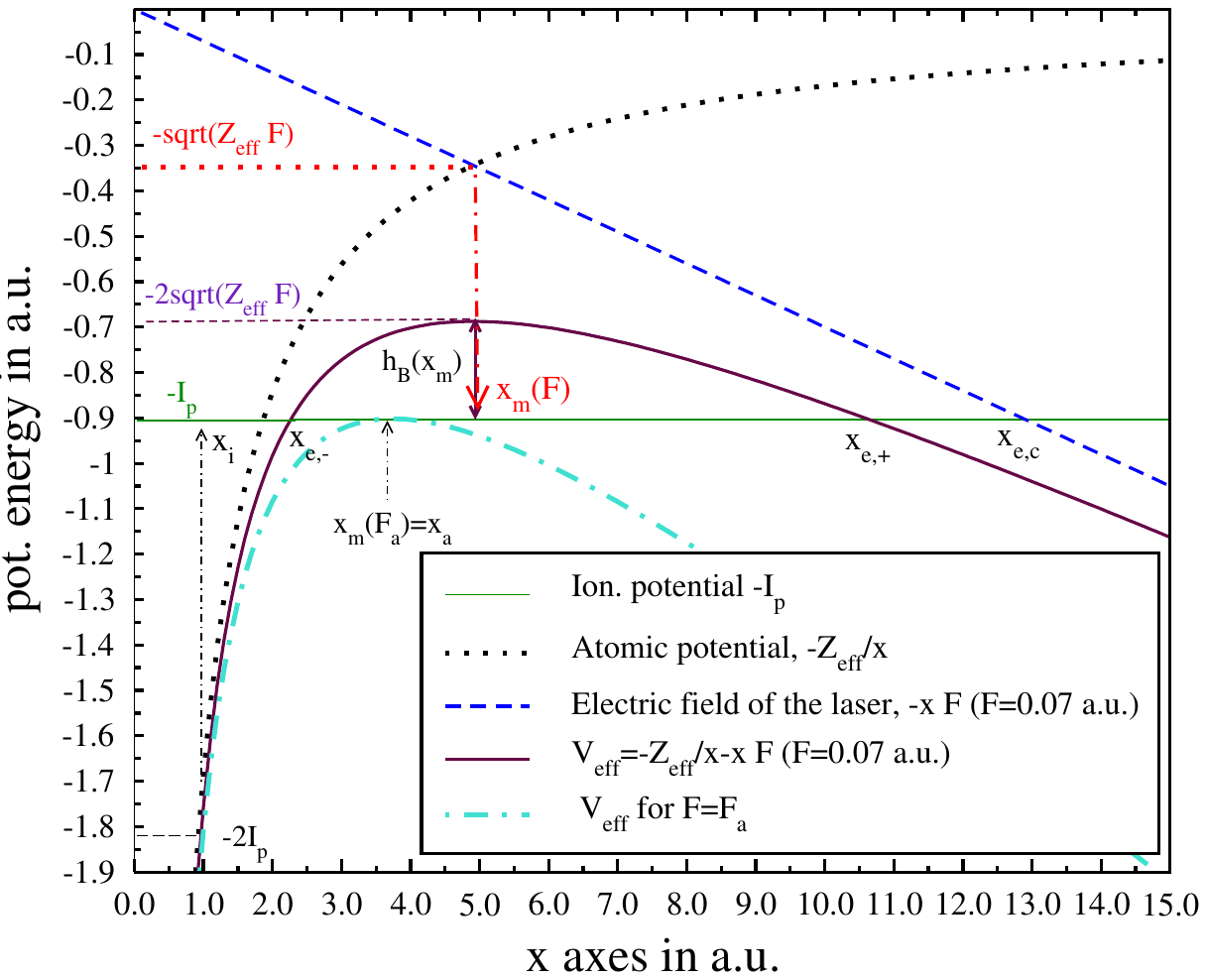}
\vspace{-7.0cm}
\caption{\label{fig:ptc}\tiny 
Graphic display of the potential curves, the barrier width and the two inner 
and outer points $x_{e,\pm}={(I_p\pm\delta_z)/2F}$, the ''classical exit`` point 
$x_{e,c}=I_p/F$ and the $x_m(F)=\sqrt{Z_{_{eff}/F}}$  the position 
at maximum of the barrier height, (note $x_a=x_m(F=F_a)$) see text.}
\vspace{-0.30cm}
\end{figure}
$\omega$ is the laser circular frequency and $F$ is the peak electric 
field strength at maximum. Here and  below we use atomic units where $\hbar=m=e=1$, 
the Planck constant, the electron mass and the unit charge are all set to 1.
At values $\gamma_{_K}>1$ one expects predominantly photo-ionization, 
while at $\gamma_{_K}<1$ ionization happens by the tunneling process, 
which means that the electron does not have enough energy to ionize but 
tunnels through a barrier made by the Coulomb potential and the electric field 
of the laser pulse to escape at the exit point to the continuum, as shown in 
fig \ref{fig:ptc}.
 
The tunneling time (T-time) in attosecond experiment is usually thought, in a simplified 
picture, as the time spent to pass the barrier region, i.e. the classically forbidden region, 
and escape at the exit point of the barrier (denoted $\tau_{_{T,d}}$) \cite{Kullie:2015I}, 
and the (quantum) particle (an electron) undergoing this process spends a time that is, 
the time needed from the moment of entering the barrier region to the moment of escaping 
the barrier region, in the tunneling direction. This picture, despite its simplicity, 
is common  between theory and experiment \cite{Kullie:2015I},  see sec \ref{ssec:CompE}.  
Moreover, in \cite{Kullie:2015I} we have defined a time-interval needed to reach 
the entrance of the barrier (denoted $\tau_{_{T,i}}$), after it is shaken off 
by the laser field at its initial position $x_i$. 
$\tau_{_{T,d}}$ is also similar to the traversal time used in context of the 
tunneling approaches \cite{Hauge:1989,Landauer:1994} such as the Feynman path 
integral (FPI) approach 
\cite{Yamada:2004,Sokolovski:1994,Sokolovski:1990,Landsman:2015} 
(and \cite{Muga:2008} chap. 7). 
But in contrast to FPI, we do not make any assumption about paths inside the 
barrier, where as well-known, the FPI approach is based on all paths starting 
at the entrance of the barrier at $t=0$  and end at the exit of the barrier at 
time $t=\tau$, which defines a time duration $\tau$. 
A second type of T-time we invented is what we call the symmetrical T-time or 
the total T-time (denoted $\tau_{_{T,sym}}=\tau_{_{T,i}}+\tau_{_{T,d}}$), where we 
found that it can be easily calculated from the symmetry property of the T-time.
$\tau_{_{T,sym}}$ is the time accounted from the moment of starting the interaction  
process, where the electron gets a shake-off, responds and jumps up to 
the tunneling ``entrance`` point taking the (opposite) orientation of the field, 
then pass the barrier region to the ''exit'' point and escapes 
the barrier to the continuum.
\vspace*{-0.50cm}                                                      
\subsection{Theory and model}\label{ssec:th}\vspace{-0.3cm}
Our start is a model of Augst et al. \cite{Augst:1989,Augst:1991}, 
where the appearance (or threshold) intensity of a laser pulse for the ionization 
of the noble gases is predicted. 
The appearance intensity is defined \cite{Augst:1989} as the intensity at which 
a small number of ions is produced.  In this model (in atomic units) 
the effective potential of the atom-laser system is given  by
\begin{equation}\label{Vx}
V_{eff}(x)=V(x)-x F =-\frac{Z_{eff}}{x}-x F,
\end{equation}
where $F=F_m$ is the field strength at maximum of the laser pulse 
 (in this work in all our formulas $F$ stands for $F_m$), and $Z_{eff}$ is the
effective nuclear charge that can be found by treating the (active) electron 
orbital as hydrogen-like, similar to the well-known single-active-electron  
(SAE) model \cite{Muller:1999,Tong:2005}.   
The choice of $Z_{eff}$ is easily recognized for many-electron systems and
 well-known in atomic, molecular  and  plasma physics 
\cite{Augst:1991,Schlueter:1987,Lange:1992,Dreissigacker:2013}. 
 We take a one-dimensional model along the x-axis as justified by Klaiber and 
Yakaboylu et al \cite{Klaiber:2013,Yakaboylu:2013}.
Indeed the barrier height at a position $x$ is, compare fig \ref{fig:ptc}:
\begin{equation}\label{hbx}
\overline{h_B(x)}=\mid h_B(x)\mid=\mid E-V_{_{eff}}(x)\mid=
\mid- I_p+\frac{Z_{eff}}{x}+x F\mid
\end{equation}
that is equal to the difference between the ionization potential and 
effective potential $V_{eff}(x)$ of the system (atom+laser) at the position $x$, 
where $E=-I_p$ is the binding energy of the electron before interacting 
with the laser.
The crossing points $x_{e,\pm}$ of $V_{_{eff}}(x)$ with the $-I_p$-line are given 
by $h_B(x)=0$
\begin{eqnarray}\label{xepm}
x_{e,\pm}=\frac{I_p\pm\delta_z}{2 F}
\Rightarrow d_B=x_{e,+}-x_{e,-}=\frac{\delta_z}{F}, \\\nonumber 
x_{e,c}=x_{e,+}+x_{e,-}=\frac{I_p}{F}=d_c
\end{eqnarray} 
where $d_B$ is the barrier width and $x_{e,c}$ is usually called the 
``classical exit'' point, it is the intersection of the field line $-x F$ 
with $I_p$-line, which equals what is usually called the ``classical'' 
barrier width $d_c$. 
$\delta_z=\delta_z(F)$ is given in eq (\ref{FFa}), we emphasize the dependence 
of $\delta_z$ on $Z_{eff}$ as done by \cite{Augst:1991,Dreissigacker:2013}. 
Augst et al \cite{Augst:1989} calculated the position of the barrier maximum 
$x_{m}$, from $\partial{V_{eff}(x)}/\partial{x}=0$
(compare fig \ref{fig:ptc}, the lower green curve) and obtained an expression 
for the atomic field strength $F_a$, 
\begin{equation}\label{Fa}
-\frac{Z_{eff}}{x_{a}}-x_{a} F =-I_{p}
\Rightarrow F_a=\frac{I_{p}^{2}}{4 Z_{eff}}
\end{equation}
and the appearance intensity $I_a=F^2_a$. 

Indeed, we can get $x_m$ and the maximum $h_B(x_m)$ from the 
derivative of eq (\ref{hbx}), ${\partial h_B(x)}/{\partial x}=0$.   
The maximum of the barrier height $h_B(x_m)$ for arbitrary field strength lies at  
$\, x_m(F)=\sqrt{Z_{eff}/F}$ \cite{Kullie:2015I}. 
Fortunately,  eq (\ref{Fa}) can be generalized as the following, for a field 
strength $F\le F_a$: 
\begin{equation}\label{FFa}
F\le\frac{I_{p}^{2}}{4 Z_{eff}}\Rightarrow \delta_z^2= I_p^2-4 Z_{eff} F\ge 0
\end{equation}
The equality $\delta_z=0$ is valid for $F=F_a$.  Indeed, 
$\delta_z =\delta_z(F)=\sqrt{I_p^2-4 Z_{eff} F}$ is a key  
quantity, it controls the tunneling process, and determines the time 
''delay'' under the barrier $\tau_{_{T,d}}$ and the total or the symmetrical 
T-time $\tau_{_{T,sym}}$, as we will see in sec. \ref{ssec:tunt}. 
The maximum barrier height $h_B(x_m)$, see fig \ref{fig:ptc}, is 
\begin{equation}\label{hbm}
h_B(x_m)=\mid\!-I_p-V_{eff}(x_m)\!\mid=
\mid\!-I_p+\sqrt{4 Z_{eff}F}\!\mid 
\end{equation}
by setting $h_B(x_m)=0$ one obtains $F_a=I_p^2/(4 Z_{eff})$, which is equivalent 
to setting $V_{eff}(x_m)=-Ip$ as done by Augst et el \cite{Augst:1989,Augst:1991}, 
and can be easily verified from eqs (\ref{hbx}), (\ref{Fa}).
\vspace*{-0.50cm}
\subsection{Tunneling time}\label{ssec:tunt}
\vspace*{-0.250cm}
In \cite{Kullie:2015I} we showed that the uncertainty in the energy can be  
quantitatively discerned from the atomic potential energy at the exit point
$\Delta E \sim \mid\! V(x_e)\!\mid=\mid\!-\frac{Z_{eff}}{x_e}\!\mid$
for arbitrary subatomic field strength $F\le F_a$. 
Then we drew the attention to the particular importance of 
the symmetry property of time, for example Aharonov-Bohm defined a time operator 
\cite{Aharonov:1961,Paul:1962,Allcock:1969I} for  a free particle.   
$\hat T= \frac{1}{2} (\hat{x} \, \hat{p}^{-1} + \hat{p}^{-1} \hat{x})$, whereas   
Olkhovsky and Recami presented 
\cite{Olkhovsky:2014,*Olkhovsky:2009,*Olkhovsky:2004,*Olkhovsky:1970} 
a more elaborate treatment (the so-called bilinear form), see also Allcock \cite{Allcock:1969I}. 
Such operators (given by Aharonov-Bohm or Olkhovsky) have the property of being 
maximally symmetric in the case of the continuous energy spectra, and the 
property of quasi-self-adjoint operators in the case of the discrete energy
spectra \cite{Olkhovsky:2009}, and are the closest 
best thing to a self-adjoined operator and satisfy the conjugate relation with 
the Hamiltonian and therefore implies an ordinary TEUR \cite{Olkhovsky:2009}  
and \cite{Muga:2008} (chap. 1). 
We used this property in the tunneling process, and we obtained 
what we called the symmetrical T-time given by \cite{Kullie:2015I}: 
\begin{eqnarray}\label{Tsym}
\hspace*{-0.5cm}\tau_{_T,sym}&=&\tau_{_{T,+}}+\tau_{_{T,-}}
=\tau_{_{T,i}}+\tau_{_{T,d}}\\\nonumber
&=&\frac{1}{2}\left(\frac{1}{\Delta E^-}+\frac{1}{\Delta E^+}\right)\\\nonumber
&=&\frac{1}{2}\left(\frac{1}{(Ip+\delta_z)}
+\frac{1}{(Ip-\delta_z)}\right)=\frac{I_p}{4Z_{eff}F}
\end{eqnarray}
where we defined $\tau_{_{T,\pm}}=1/(2\Delta E^{\mp})={(2(Ip\pm\delta_z))}^{-1}$, 
or $(1/2)\Delta E^{\pm}=\Delta E(x_{e,{\pm}})=\mid\!\!-\frac{Z_{eff}}{x_{e,\pm}}\!\!\mid$,   
and $\tau_{_{T,i}}=\tau_{_{T,+}}$, $\tau_{_{T,d}}=\tau_{_{T,-}}$.
The barrier itself causes a delaying time relative to the atomic field 
strength given by: 
\begin{equation}\label{Td}
\tau_{_{T,d}}\equiv\frac{1}{2 \Delta E^+}
=\frac{1}{2\,(I_p-\delta_z)}=\tau_{_{T,-}}\,(\mbox{for } F\le F_a)
\end{equation}
which we call $\tau_{_{T,d}}$  meaning that is the time 
delay with respect to the atomic field strength (the time duration) to pass  
the barrier region (in the direction $x_{{_e,-}}\!\rightarrow\! x_{_{e,+}}$) 
and escape at the exit point $x_{e,+}$ to the continuum, 
for more details see \cite{Kullie:2015I}. 
The first term in eq  (\ref{Tsym}) is the time needed to  reach the ''entrance`` 
point $x_{e,-}$, compare fig \ref{fig:ptc}.
\begin{equation}\label{Ti}
\tau_{T,i}= \tau_{(x_i,x_{e,-})}=
\frac{1}{2\Delta E^-} = \frac{1}{2(I_p+\delta_z)}=\tau_{_{T,+}}
\end{equation}
At the limit $F\rightarrow F_a$, $\delta_z\rightarrow 0$, and the tunneling time 
becomes the ionization time, 
\begin{equation}\label{Tz}
\tau_{_T,sym}=\frac{1}{2Ip}+\frac{1}{2Ip}=\frac{1}{Ip}
\end{equation}
accordingly, the tunneling happens in two steps \cite{Kullie:2015I}, the  
shake-up $\tau_{_T,i}=\frac{1}{2Ip}$ and shake-off $\tau_{_T,d}=\frac{1}{2Ip}$.
For $F=F_a$ the two steps happen with equal time duration and are not strictly 
separated, unlike the case for subatomic field strength $F<F_a$ due to the 
barrier, as discussed in \cite{Kullie:2015I}. 
It is nice to see that the mean of the uncertainties
$\Delta E_{M}=\frac{1}{2}\left({\Delta E^-}+{\Delta E^+}\right)=I_p$, 
this has led some authors to mention that a T-time of the order $1/I_p$ can be 
estimated via TEUR,  when assuming an energy uncertainty of order of $Ip$ 
\cite{Klaiber:2013}. 

The T-time found in eq (\ref{Tsym}) can also be derived in an elegant way, when
we assume that the barrier height eq (\ref{hbm}) corresponds to a maximally 
symmetric operator as the following.  
The barrier height eq (\ref{hbm}) $h_B(x_m)\equiv h_M$ can be related to a (real) 
operator $\widehat{h}_B(x_m)$
\begin{equation}\label{hm1}
\widehat{h}_M^{\pm}=-I_p\pm\sqrt{4 Z_{eff}F}= -I_p\pm\Delta
\end{equation}
and the uncertainty in the energy caused by the barrier
\begin{equation}\label{hm2}
\left(\widehat{h}_M^{-} \widehat{h}_M^{+}\right)^{1/2}=
[(-I_p-\Delta)(-I_p+\Delta)]^{1/2}=\delta_z
\end{equation}
From this we get ${\Delta E^{\pm}}=abs(-Ip\pm\delta_z)$ 
and hence  $\tau_{_{T,d}}$, $\tau_{_{T,i}}$ and $\tau_{_{T,sym}}$  
by the virtue of eq (\ref{ucr}), where we assumed that the time is intrinsic 
and has to be considered with respect to the atomic field strength, i.e.  to 
the ionization potential $I_p$, see discussion in subsec \ref{sssec:IEt}.  
\vspace*{-0.50cm}
\subsection{Comparison to the experiment}\label{ssec:CompE}
\begin{figure}[t]  
\vspace{-3.750cm}
\includegraphics[height=13.0cm,width=11cm]{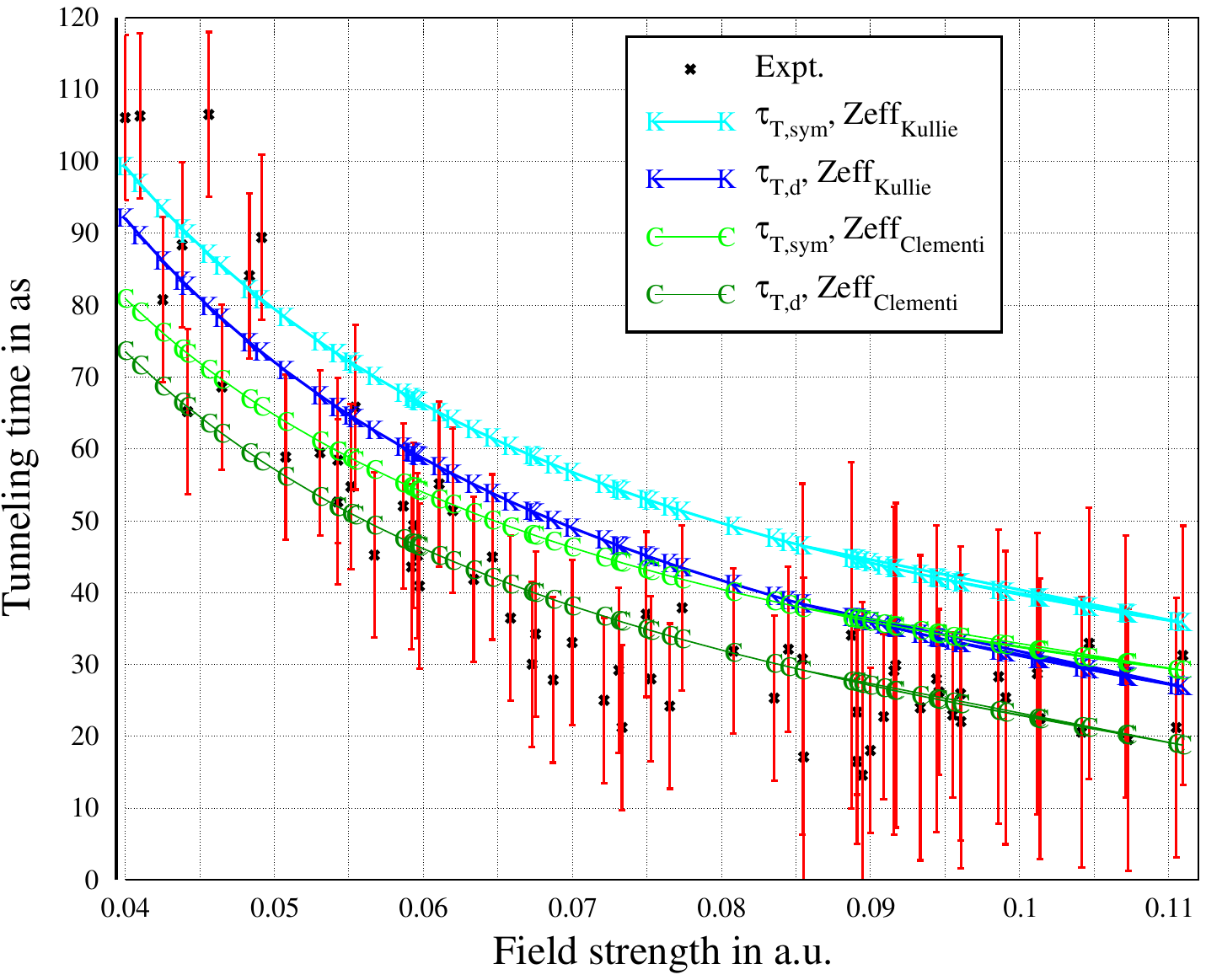}
\vspace{-5.0cm}
\caption{\label{fig:tut1}\tiny 
T-time  $\tau_{_{T,d}}$ eq (\ref{Td}) and $\tau_{_{T,sym}}$ eq (\ref{Tsym}),  
for two different $Z_{eff}$ models. 
Time is in attosecond units vs laser field strength in atomic
units, corresponds to the tunneling ionization of the He-Atom in strong field,
in good agreement with experimental result
\cite{Landsman:2014I,Landsman:2014II,Eckle:2008s,Eckle:2008}.
Experimental values are kindly sent by A. S. Landsman \cite{Landsman:2014II}.}
\end{figure}

\vspace*{-0.25cm}
In fig \ref{fig:tut1} we show the results for $\tau_{_{T,d}}$ eq (\ref{Td}), 
and the symmetrical 
(or total) T-time $\tau_{_{T,sym}}$ of eq (\ref{Tsym}). 
Note eq (\ref{Td}), is the second term of eq (\ref{Tsym}), which is the actual 
T-time, i.e. the time needed to pass the barrier region 
$(x_{e,-}\rightarrow x_{e,+})$ and escape at the exit point to the continuum. 
The plotted result is for the He-atom together with the experimental result of 
\cite{Eckle:2008s,*Eckle:2008}. 
The experimental data and the error bars in the figure were kindly sent by 
A. S. Landsman \cite{Landsman:2014II}.   
We plotted the relations 
at the values of the field strength at the maximum of the elliptically polarized 
laser pulse ($\lambda =735$, elliptical parameter  
$\varepsilon=0.87$, $F=F_0/\sqrt{1+\varepsilon^2}$) used by the experiment 
exactly as given in \cite{Landsman:2014II}. 
 Concerning the experiment by the group of Keller et al 
\cite{Landsman:2014II}, the time delay due the barrier is measured indirectly 
$t_T^{exp}=\frac{(\theta_m-\theta_C)}{\omega}$, where $\omega$ is the laser 
frequency, $\theta_m$ the angular offset of the center of angular distribution, 
and $\theta_C$ a correction due to the Coulomb force of the ion calculated by 
classical trajectory simulation  
\cite{Eckle:2008s,Eckle:2008,Landsman:2014II,Landsman:2015,Hofmann:2013}. 
Unfortunately, in this experiment the beginning of the interaction between 
the laser field and the bound electron cannot be directly observed or exactly 
determined.
The time zero $t_0^{exp}$ is one of the assumption of the model, which used to 
interpret the data. Consequently, in the experiment it is not possible to 
distinguish between the instant of the interaction $t_0=0$ and the instant when 
the electron enters the barrier region, say $t_{T,l}$, \cite{Hofmann:2015}.
However, arguing that the probability of tunneling is highest when the barrier 
is shortest, corresponding to $t_0^{exp}$ \cite{Hofmann:2015}, 
suggests that the data from the measurement are comparable to the time delay   
caused by the barrier (the time spent in the classical forbidden region 
\cite{Hofmann:2015}), which means that $t_T^{exp}$ corresponds or 
(approximately) equals the actual T-time delay of our model 
$t_T^{exp}\approx\tau_{T,d}$. 

The two different effective charge models are from Kullie \cite{Kullie:1997},  
with $Z_{_{eff,K}}=1.375$  and Clementi \cite{Clementi:1963} with 
$Z_{_{eff,C}}=1.6875$.
For the $\tau_{_{T,d}}$ we see an excellent agreement with the 
experiment. $\tau_{_{T,d}}$ corresponds to the T-time measured in the experiment, 
that is the time (interval) needed to pass the barrier region from  the entrance 
to the exit point and escape to the continuum with a shake-off \cite{Kullie:2015I}, 
or between the instant of orientation at $x_{e,-}$ and the instant of ionization at 
$x_{e,+}$, which is the time spent in the classically forbidden region 
\cite{Eckle:2008s}.   

In fig \ref{fig:tut1} we see that the difference between the total or symmetrical 
T-time $\tau_{_{T,sym}}$ and the (actual) T-time $\tau_{_{T,d}}$ is small, 
because the second term  $\tau_{_{T,d}}$ in eq (\ref{Tsym}) incorporates the 
delay time caused by the barrier, and is the main time contribution to the 
tunneling process for a large barrier. 
Whereas the first part term  $\tau_{_{T,i}}$, is due to the shake-up 
of the electron by the field moving it to the ''entrance'' $x_{e,-}$, 
which is small for a small $F$. 
For a large field strength the two parts become closer because the barrier width 
is getting smaller $\delta_z/F=(x_{e,+}-x_{e,-}) \rightarrow 0$ and for the 
appearance intensity ($\delta_z =0)$ they are equal.  
For small $F\lesssim 0.055\, au$,  $Z_{eff,K}$ 
gives a better agreement with the experiment, whereas for larger field strength 
$Z_{eff,C}$ is more reliable, where multielectron effects are expected due to 
the decreasing width of the barrier and the tunneling electron is closer to the 
first one, when it traverses the barrier.

At the limit $F=F_a$ of the sub-atomic field strength $F<F_a$, the 
tunneling process is out and an ionization process called 
''above the barrier decay`` is beginning. 
For super-atomic field strength $F>F_a$, $\delta_z$ becomes imaginary (and so 
the crossing points, compare eq (\ref{xepm})), which indicates that the real 
part $\frac{1}{2I_p}$ of $\tau_{_{T,d}}$ or $\tau_{_{T,i}}$, is the limit for 
a ''real'' (see below) time (tunnel-)ionization process.  
Indeed, in this case the atomic potential is heavily disturbed and the imaginary 
part of the time $\tau_{_{T,d}}$ is obviously, due to the ionization or the 
release of the electron (to the continuum) from a lower level than $-Ip$ 
(and possibly escaping with a high velocity), where the ionization happens mainly 
by a shake-off step \cite{Delone:2000} (chap. 9).  
Here we see the clear difference between the quantum mechanical and the classical 
clocks \cite{Aharonov:1961}. 
Classically we can make the interaction time with the system arbitrarily small,  
the real part of the time can be made arbitrary small, and an imaginary part 
is absent. 
In quantum mechanics the tunneling-ionization time has a real part 
limit $\tau_{_{T,d}}=1/(2I_p)$, an imaginary part arises when the field strength 
is larger than the atomic field strength $F_a$, in both terms $\tau_{_{T,i}}$ 
and $\tau_{_{T,d}}$.
 
However, in our treatment, although, $\tau_{_{T,i}}, \tau_{_{T,d}}$ have  
an imaginary part for $F\ge F_a$, the total or symmetrical T-time remains real 
\[\tau_{{_T,sym}}=\frac{I_p}{4 Z_{eff} F}=\frac{1}{I_p \xi}=
\frac{1}{I_p}(\frac{F_a}{F})\] 
for ionization processes with an arbitrary field strength, where $\xi=\frac{F}{F_a}$ 
is the perturbation parameter, for which the perturbation theory is valid, 
when $\xi\ll1$. 
$\tau_{_{T,sym}}$ is small for $\xi=\frac{F}{F_a}\rightarrow c, c\ge1$, 
and probably the above relation ($\tau_{_{T,sym}}$) loses its validity for $F>F_a$,  
suspecting a break of some symmetry at $F=F_a$,  non-linear effects arise and 
the interaction becomes physically a different character. 
It is apparent from $\tau_{{_T,sym}}$ eq (\ref{Tsym}), that 
the T-time has no imaginary part, when the symmetry of the time is considered, 
i.e. assuming the maximally symmetric (quasi-self-adjoint) 
property discussed in detail by Olkhovsky \cite{Olkhovsky:2009}. 
\vspace{-1.0cm}
\subsection{Tunneling, Mandelstam-Tamm and Keldysh time}\label{ssec:TMK}
\vspace{-0.25cm}
\vspace{-0.50cm}
\subsubsection{Preliminary discussion}\label{sssec:prlim}
\vspace{-0.250cm}
An important point in our treatment (T-time approach) is that, 
our T-time is intrinsic or dynamically connected to the system and is the T-time of 
a quantum particle, whereas for example the Keldysh time (K-time) $\tau_K$ is defined 
for a classical particle, and $\tau_K\!\!=\!\!{\sqrt{2I_p}}/{F}$ is determined as 
\cite{Orlando:2014I,*Orlando:2014II}
\begin{equation}\label{tkev}
\overline{\tau}_K\!\!=\!\!d_c/|v|
\end{equation} where $d_c\!\!=\!\!I_p/F$ is the ``classical'' barrier width 
(compare with $d_B$ eq (\ref{xepm})) and $|v|$ is the average speed of an electron 
under the static barrier. 
We note that in this definition $d_c$ is equal to the ''classical`` exit point 
$x_{e,c}$ eq (\ref{xepm}), compare fig \ref{fig:ptc}. 
It is easy to find that the average speed $|v|$ is the (arithmetic) mean of the 
speed $\overline{v}=(v_e+v_0)/2\!=\!{(0+v_0)}/2\!=\!\sqrt{2I_p}/2$,
neglecting the atomic potential, where $v_e=0$ is the velocity at the exit point 
according to the strong field approximation (SFA). 
$\overline{v}$ equals the ''Wirkungswellen-'' velocity (action-wave velocity) $u$ 
of a classical particle with a total energy equal  to the kinetic energy 
$E=T, V=0$, $u=\frac{|E|}{\sqrt{2 T}}\!=\frac{\!\sqrt{2I_p}}{2}$ 
\cite{Nolting:1990} (chap 3.6, p. 196).
Then, $\tau_K=d/\overline{v}=(I_p/F)\cdot (2/\sqrt{2I_p})=\sqrt{2I_p}/F$ 
and the electron tunnels like a classical particle with $\overline{v}=u$ 
through the barrier (neglecting the potential $V=0$). 
Hence our T-time $\tau_{T,d}$ (and $\tau_{T,\pm}, \tau_{T,sym}$) represents 
a quantum clock, whereas $\tau_K=\overline{\tau}_K$ represents a laboratory or 
external time, it cannot describe the attosecond experiment and contradicts 
the TEUR as mentioned by Rzazewski \cite{Rzazewski:1993}. 

Orlando et al \cite{Orlando:2014I,*Orlando:2014II} calculated the Mandelstam-Tamm time 
(MT-time, $\tau_{MT}$) and the result equals the K-time 
$\tau_{MT}\!\!=\!\!\tau_K=\overline{\tau}_K$.  
 They assumed that the uncertainty in the energy due to the barrier is given by  
the standard deviation $\delta E\!\!=\!\!\Delta H\!\!=\!\!\sqrt{<H^2>-<H>^2}$, 
where $H=H_0- x F$ and $H_0$ is the Hamiltonian of the unperturbed atomic system
(and $(\delta p)^2\sim2<\!\!T\!\!>=2I_p$) \cite{Orlando:2014I,*Orlando:2014II}. First, it is known 
that the standard deviation can be a very unreasonable measure for the uncertainty 
in the energy \cite{Uffink:1993} especially for wave functions with long tails. 
Secondly, $\Delta H$ will not lead to a correct T-time, because the uncertainty in 
the energy is not determined by $\Delta H$ but by the barrier height given by 
eqs (\ref{hbm}), (\ref{hm1}-\ref{hm2}), which, at the end, leads to $\Delta E^{\pm}$ 
and the correct T-time $\tau_{_{T,\pm}}$. Note in our case the height and the 
width of the barrier are not independent,  both depend on the filed strength, 
see eqs (\ref{hbm}), (\ref{xepm}) .  
However, it is somehow puzzling to see that our 
$\delta E=abs(-I_p+\delta_z)=(Ip-\delta_z)=\Delta E^+$ is the expectation value 
of the Hamiltonian  $H_B=H_0+(x_{e,+}-x_{e,-})F=(-Ip+\delta_z)$, 
where $d_B=(x_{e,+}-x_{e,-})=\delta_z/F\ne d_c$ is 
the barrier width (compare eq (\ref{xepm})), this leads to the correct T-time 
eq (\ref{Td}) by the virtue of eq (\ref{ucr}). 
From this, it is also true that the time should be considered from within the 
system (internal time), where the reference point is (the ionization potential) 
$abs(E_0)=I_p$, which is overcome when $F=F_a$, and the system undergoes  
an (ejection of an electron or a threshold) ionization process (no barrier). 
The delay time is found relative to the reference point from the perturbation 
$d_B\cdot F=(\frac{\delta_z}{F})F=\delta_z$ as we already  discussed,
which becomes zero at atomic field strength, $\delta_z(F_a)=0$ (no barrier), 
and the T-time $\tau_{T,d}$ is reduced to $1/(2I_p)$, the second part of the 
total ionization time, see eq (\ref{Tz}).
   
A further point, Orlando et al \cite{Orlando:2014I,*Orlando:2014II} obtained a similar result by 
numerical investigation of the wave packet dynamics with the same Hamiltonian $H$, 
which in turn, makes the puzzle more difficult to understand. 
Nevertheless, most important is the preparation of the system and 
its Hamiltonian and how to observe the time in the system, see discussion further  
in subsec \ref{sssec:IEt} and subsec \ref{sssec:what}. 
In fact Maquet et al \cite{Maquet:2014} noted that a time-delay  requires to 
choose a reference system, delays in numerical simulation can refer in principle 
to any arbitrarily chosen reference \cite{Maquet:2014}, 
see next subsec \ref{sssec:IEt}.   
\vspace{-0.50cm}
\subsubsection{Internal- and external- time frame}\label{sssec:IEt}
\vspace{-0.250cm}
Indeed, we can follow the procedure of Orlando et al but with two 
crucial differences.
The first difference is, we take $<x^2> =d_B^2$, which means that the particle 
is de-localized over the whole barrier instead of 
$1/(\Delta p)^2,\, (\Delta p)^2=2<\!\!T\!\!>$ used by Orlando  et al 
\cite{Orlando:2014I,*Orlando:2014II}, this assumption is quantum-mechanically valid.
Calculating now the ''variance'' as done in \cite{Orlando:2014I,*Orlando:2014II}, we get  
$\delta E\!\!=\left((I_p^2-d_b^2\cdot F^2)-I_p^2\right)^{1/2}= \delta_z$.
The second difference, we have to take our uncertainty in the energy (and thus the T-time) 
relative to the binding energy in the ground state $-Ip$, which is decisive 
(see below and compare fig \ref{fig:ptc}), thus we obtain $(-Ip+\delta)$ 
(or $(-Ip\pm\delta)$ considering the symmetry), 
this leads to the ''correct'' T-time 
$\tau_{T,d}$ (or $\tau_{T,\pm}=(-Ip\mp\delta)^{-1}$ and $\tau_{T,sym}$).

In fact, one can argue that $-Ip$ must be taken to avoid the divergence of 
the time to infinity for $F=F_a$, because it is physically incorrect, as 
$\delta_z(F_a)=0$, which in turn can be seen as an initialization of the internal 
clock, i.e. the T-time is counted as a delay with respect to the ionization 
at $F_a$ (the limit of the subatomic field strength). 
This is exactly what Maquet et al \cite{Maquet:2014} alluded to, quoting them 
``the definition of a ‘time-delay’ requires to choose a ‘reference’ system.''
The other limit, i.e. for $F\rightarrow 0,\, (I_p-\delta)\rightarrow0$ is then 
the multiphoton regime, the tunneling did not take place and for 
$F=0, \delta_z=I_p\Rightarrow \tau_{T,d}=\infty$, hence nothing happens. 
That is exactly the well-known limit of an energy eigenstate of the 
MT-time relation, where the rate of change of a property 
decreases with increasing sharpness of the prepared energy 
$(-I_p+\delta_z(F\rightarrow0))\rightarrow0$.    

Furthermore, we can consider the barrier width as the sum of the distances to 
reach, and from  the ionization exit $x_a$ (at the atomic field strength), 
then (compare fig \ref{fig:ptc})
\[d_B=\delta_z/F=x_{e,+}-x_{e,-}=(x_a-x_{e,-})+(x_{e,+}-x_a)\]     
which causes an energy ``uncertainty`` equal to 
$abs(-I_p+\delta_z)=abs(-I_p+(\delta_z/F)F)=abs(-I_p+d_B F)$. It can 
be considered as the distance to the ionization limit ($d_B=0, \delta_z=0$) 
on the energy scale, and hence the time delaying $\tau_{T,d}$ to pass the barrier region. 
The point is, one would say that the two points $x_{e,\pm}$ are similar to  
the two slits (in the double slit experiment) with a distance $d_B$ 
(between the slits).
\begin{figure}[t]  
\vspace{-4.10cm}
\hspace*{-0.250cm}\includegraphics[height=13.225cm,width=11.cm]{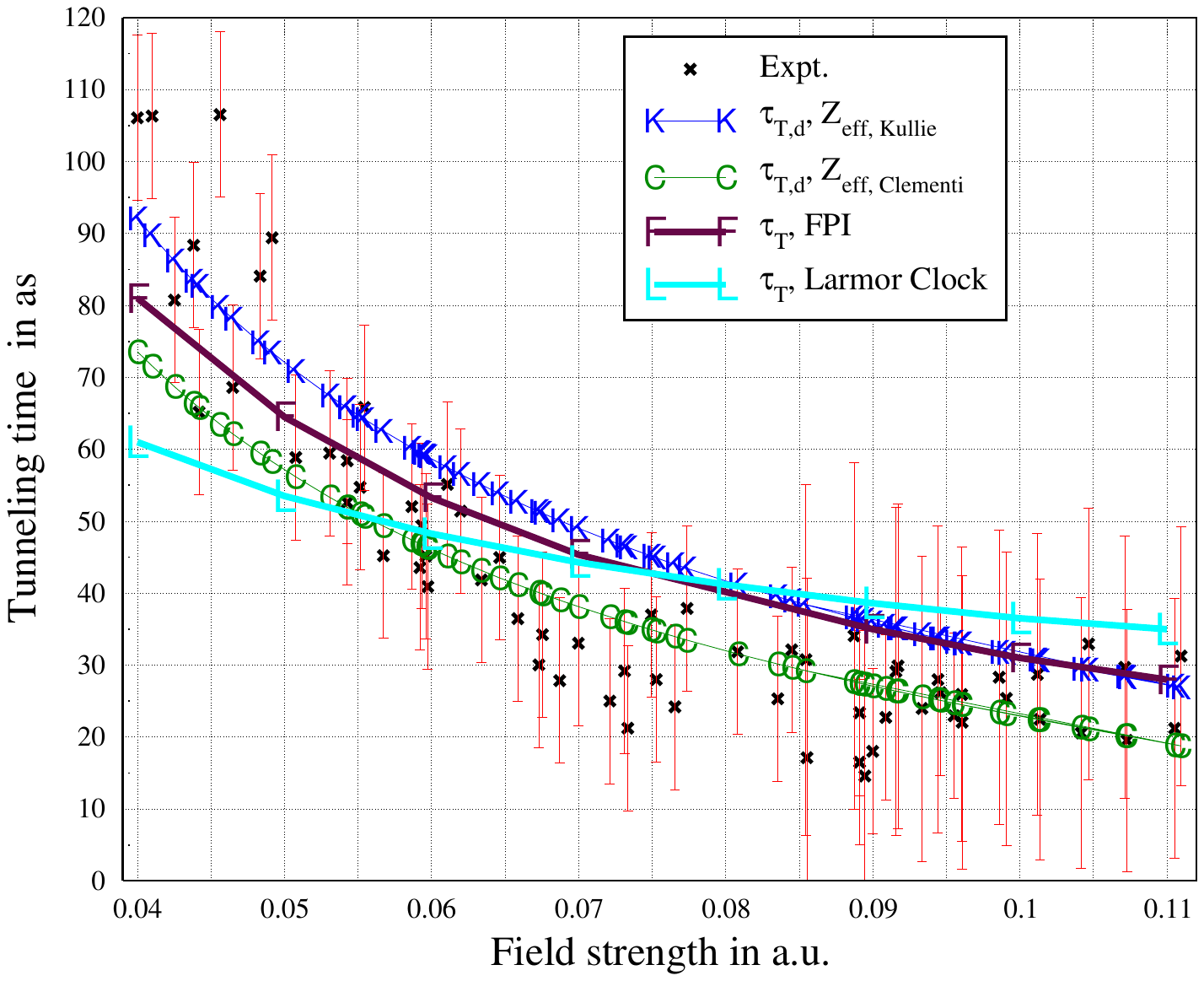}
\vspace{-5.cm}
\caption{\label{fig:FPI}\tiny 
T-time  $\tau_{_{T,d}}$ eq (\ref{Td}) for two different $Z_{eff}$ models as 
in fig \ref{fig:tut1} together with the FPI and Larmer Clock results 
\cite{Landsman:2014II}, and the experimental result \cite{Eckle:2008s,Eckle:2008,Landsman:2014II}.
Experimental values,  the FPI and Larmer result are kindly sent by A. S. Landsman 
and C. Hofmann \cite{Landsman:2014II,Hofmann:2013}.} 
\end{figure}
Accordingly, in the Feynman path integral formalism, the dwell time or the 
residence time (which is the time a particle spends in the barrier region),  
can be seen as a result of the interference between the wave packets or paths. 
Forwards (to $x_{e,+}$) and backwards (to $x_{e,-}$) tunneling correspond 
to the transmission and reflection amplitudes of the wave packet, respectively. 
This picture strengthens the good agreement between our result and FPI  
shown fig \ref{fig:FPI}.

For the experimental setup the internal time view is not a crucial point, since 
in the experiment one extracts the internal information of the system from the 
output on the screen, whereas from a theoretical point of view it is very crucial, 
and a controversial discussion about the T-time is continuing. 
As already mentioned above Maquet et al pointed out that 
the definition of a ‘time-delay’ requires to choose a ‘reference’ system. 
Delays in ‘Gedanken’ experiments or in numerical simulations can refer in principle
to any arbitrarily chosen reference \cite{Maquet:2014}. 
The situation is similar to what occurs in the special relativity, where a moving 
particle has its own time in its inertial frame, which differs from the time of 
the viewpoint of another inertial frame. 
A well-known example in this context, is the Muon decay, which can be found 
in introductory books.
About $10000$ Muons reach every square meter of the Earth's surface per minute.  
Although their lifetime without relativistic effects would allow a half-survival 
distance of only about $456 m$ at most (as seen from Earth).  
The time dilation effect of special relativity (from the viewpoint of the Earth)
allows cosmic ray secondary Muons to survive the flight to the Earth's surface,  
since in the Earth frame, the Muons have a longer half lifetime due to their high 
velocity $0,9997\,c$, where $c$ is the speed of light.  
From the viewpoint (inertial frame) of the Muon, on the other hand, it is the 
length contraction effect of special relativity which allows this penetration, 
since in the Muon frame, its lifetime is unaffected, but the length contraction 
causes distances through the atmosphere and Earth to be far shorter than these 
distances in the Earth rest-frame. 
In our case of the T-time, it is not the relativistic effect, but observing   
the time from within the system (internal clock \cite{Aharonov:2000}), which  
differentiates it form the classical view. 
And although the relativistic effects in this case are negligible, the special 
relativity point of view is helpful, first to understand our current case and 
second, it could be necessarily (by using Dirac instead of SEQ) 
to establish the comparability of space and time on equal footing, where the 
symmetry of the time is then naturally embedded in the theory 
\cite{Bauer:2014,Bauer:1983,Antoniou:1992,Prugovecki:1982}. 
However, Dodonov \cite{Dodonov:2015} claims that no unambiguous and generally 
accepted results have been obtained so far. 
\vspace{-0.5cm}
\subsubsection{Critique of the Keldysh time and beyond}
\label{sssec:crit} 
\vspace{-0.25cm}
We discuss here some points about the K-time and MT-time, with the hope, that they 
are hints in understanding the puzzle of the T-time in attosecond experiment and 
the different views to calculate it.  

First, one of the failures of K-time comes form its inadequate definition. 
Then,  the difference between $d_c$ and $d_B$ is large (compare eq (\ref{xepm}) 
and fig \ref{fig:ptc})  $d_B<d_c, d_c-d_B=(I_p-\delta_z)/F=2 x_{e,-}$,  
and although 
\[\lim_{F\to 0} d_B=\lim_{F\to 0}\frac{\delta_z}{F}=
\lim_{\xi\to 0}\frac{I_p}{F}\sqrt{1-\xi}=\frac{I_p}{F}=d_c\] 
but for $F\rightarrow 0, (\xi=\frac{F}{F_a}\ll1, \mbox{the perturbation regime})$,  
for a fixed frequency $\omega$, one obtains 
$\gamma_K(F\rightarrow 0)=\omega\cdot\tau_K(F\rightarrow 0)\, \gg\,1$ 
and that is  the region of photo-absorption \cite{Delone:2000} and tunneling is 
less probable compared with the multiphoton processing,   
because the K-time is based on  the competition with the absorption of quanta  
\cite{Rzazewski:1993}. Thus the tunneling is unlikely to happen 
at a (real) time interval length equal to K-time for small $F$. 
Moreover, for stronger field ($F\rightarrow F_a$) $\delta_z$ is small, hence 
$d_B(F\rightarrow F_a) \approx 0$ is much smaller than $d_c$ 
$(\lim_{\xi\to 1}\frac{d_B}{d_c}=\lim_{\xi\to 1}\sqrt{1-\xi}=0)$. 
For $F=F_a, d_B= 0$, whereas 
$d_c=I_p/F_a=\frac{I_p}{Ip^2/4Z_{eff}}=4Z_{eff}/I_p$.
Which means that the definition of the ``classical`` barrier width as usually 
done to calculate the K-time, is inaccurate and inadequate. 

And second, it is interesting to show that the MT-time 
$\tau_{MT}$ can be directly obtained. 
Then, $\tau_{MT}$ can be obtained by using the momentum $\hat p$ of the 
tunneling-electron as a dynamical observable, $\tau_{MT}$ results from its 
definition (again neglecting the atomic potential), using $(\Delta p)^2=2<T>$ 
as done by Orlando et al \cite{Orlando:2014I,*Orlando:2014II} we get:
\begin{equation}\label{tMT}
 \tau_{MT}^{p}=\frac{\Delta p}{\mid\frac{d}{dt}<p>\mid}=
\frac{\sqrt{2I_p}}{F}=\tau_{MT}
\end{equation}
In fact, this is the classical definition of the time a particle spends, 
when it moves rectilinear with a deceleration $F$ and an initial/final momentum 
$p_0/p_f$ between the initial and final moment $t_{0}, t_{f}$ respectively:    
\[\tau^{cl}=t_{f}-t_{0}=\frac{\Delta p}{\frac{d}{dt}p}=\frac{\Delta p}{F}=
\frac{\sqrt{2I_p}}{F},
\quad \mbox{with } \Delta p=\mid\!p_{f}-p_{0}\!\mid\] 
Obviously, from $\tau^{cl}=\tau_{MT}^{p}=\tau_{MT}=\tau_{K}=\overline{\tau}_K$ 
(compare eq. (\ref{tkev})), it follows that the MT-time derived this way by Orlando 
et al (among others), which is equal to the K-time, is not the tunneling time of 
a quantum particle, its relation with the TEUR  and its connection to K-time still 
need to be explained, for this task we discuss it further below 
(subsec \ref{sssec:what}).  
But again quantum-mechanically (see above and eq (\ref{Tsym})-(\ref{Ti})) with 
$\Delta p=1/\sqrt{<x^2>}=1/d_B=\frac{F}{\delta}$ and considering the symmetry 
and the initialization of the internal clock as discussed above, one obtains 
the T-time (or the delay time) 
$\tau_{T,d}$ and $\tau_{T,i}, \, \tau_{T,sym}$:
\[\frac{\Delta p}{\mid\frac{d}{dt}<p>\mid}=
(\frac{F}{\delta})\frac{1}{F}=\frac{1}{\delta_z}\Rightarrow\tau_{\pm}
=\frac{1}{2(I_p\pm\delta_z)}\] 
This derivation seems somehow plausible, nevertheless its validity is obvious 
(see eqs (\ref{Tsym})-(\ref{Td})) and shows the importance of, first the internal 
time point of view, i.e. the factor $I_p$, which means the time delay is accounted 
relative to a reference point, which is determined by the internal properties of the QM-system.
The reference point is the ionization potential, which corresponds to the ionization 
at atomic field strength, where $\tau_{\pm}=\frac{1}{2I_p}$  and 
$\tau_{T,sym}=\tau_{T,+}+\tau_{T,-}=\frac{1}{I_p}$.

Finally, it is  worth noting that,  
with the definition of Mandelstam-Tamm time \cite{Gray:2005,Sukhanov:2000} 
($ A, B $ are two operators of the system, and $\psi$ is its wave function)
\begin{eqnarray}\nonumber
\Delta_{\psi}A \Delta_{\psi} B\ge \frac{1}{2}
\mid\left< i(\widehat{A}\widehat{B}-\widehat{B}\widehat{A})(\psi),\psi\right>\mid\\\nonumber
\Delta t_A=t_{\psi,A} 
=\frac{\Delta_{\psi} A}{\mid \frac{d<A>}{dt}\mid}
\end{eqnarray}
$\tau_{MT}$ is taken to be 
$(\Delta t_A)_{min}$ the minimum of all 
possible quantities, this choice is rather arbitrary because it essentially depends 
on the state used to average this expression \cite{Sukhanov:2000}.

The authors of  \cite{Gray:2005} argued that the quantity 
$\Delta_\psi t \equiv t_A\equiv\tau_{MT}$, which occurs in the TEUR-version of 
Mandelstam-Tamm depends on  $\widehat H$ and on the states $\psi$ 
that are not eigenvectors of an observable, it is not the standard deviation of 
an observable, but is the infimum of the ratio of static uncertainty to dynamic 
uncertainty per unit time (compare eq \ref{tMT}). This certainly contradicts what 
Orlando et al claim that $\tau_K=\tau_{MT}$ represents a lower limit for the
tunneling time of an atomic system as measured by a generic quantum clock, 
apart form the fact, that $\tau_K$ represent a classical (laboratory/external) 
clock, as we are discussing in this work.  
Furthermore, the authors of  \cite{Gray:2005} argued that, although changes in the mean of a variable are 
important, we cannot argue that they are the only significant measures of change 
as time passes or that any change smaller than one standard deviation is truly 
negligible. 
Additionally, according to  Gray et al \cite{Gray:2005}, Messiah \cite{Messiah:1961} 
calls $\Delta t_\psi$ ``the characteristic time of evolution'', but they claim  
that this formulation recommends itself since it does not promise too much, 
and they provided, what they called a (mathematical) precise 
definition of $\Delta t_\psi$.

The characteristic evolution time of Messiah seem to agree with the picture 
brought by Zhao et al \cite{Zhao:2013}. 
Zhao et al studied the ionization and tunneling times in high-order harmonic 
generation. They suggested that the imaginary part of the T-time equals the K-time, 
by numerically solving the SEQ for He-atom using the imaginary 
time evolution (and quantum-mechanical retrieval method of the tunneling time based 
on trajectories evolving in complex time). 
Indeed, Zhao et al define T-time as the imaginary part of the complex ionization 
time, it was in addition related to a real (time) quantity, i.e. resulting from the 
the retrieval based on classical dynamics (using real times), or the classical 
three-step model, the result was also shown to be in agreement with 
the quantum orbit model \cite{Lewenstein:1994,Salieres:2001}. We discuss this further 
in the next subsec \ref{sssec:what}.      

\vspace{-0.5cm}
\subsubsection{Tunneling time, what does it means?}\label{sssec:what}
\vspace{-0.25cm}
So far we are not claiming to resolve the puzzle of the T-time. 
However, the situation is the following.  
Our result, fig \ref{fig:tut1} (see also \cite{Kullie:2015I}), is in excellent 
agreement with the experiment and with the Feynman path integral (FPI) and the 
Larmer clock of Landsman et al \cite{Landsman:2014I,Landsman:2014II} 
(compare fig \ref{fig:FPI}).  
In our treatment we made use of the fact that the internal  time (i.e. 
dynamically connected to the system), is a central point of the time theory in 
quantum mechanics, which differs fundamentally from the parametric 
external/laboratory time (or classical time) \cite{Busch:1990I,Busch:1990II}. 
The K-time (and the MT-time) of Orlando et al \cite{Orlando:2014I,*Orlando:2014II} 
(among others), which is assumed to be real, differs too much from the 
experimental finding \cite{Landsman:2014II,Landsman:2014I,Landsman:2015}. 
In our critique on the K-time and the MT-time, we showed many hints 
and included remarks from some authors \cite{Gray:2005,Sukhanov:2000}, 
which suggests that a K-time is inadequate to treat the (real) T-time in 
attosecond experiment.

In fact, Orlando et al assists their derivation with a numerical wave dynamics 
treatment, whereas Zhao et al \cite{Zhao:2013} et al suggested that 
the imaginary part of the T-time equals the K-time, by a retrieval 
quantum-mechanical method of the T-time based on complex-time trajectories. 
Following the Messiah idea \cite{Gray:2005,Messiah:1961} 
(MT-time as ``the characteristic time of evolution''), the MT-time can be brought 
under the same hat, the similarity is evident with the imaginary (Keldysh) 
time of Zhao et al.   

It stands to reason the idea that the K-time of Orlando et al, which is found by 
the numerical investigation of the wave function dynamics, is identical with 
(imaginary) K-time of Zhao et al, which is found by the numerical solving of 
the SEQ with imaginary time evolution, and the K-time $\tau_k$ represents 
the external-type of time. 
The real T-time measured by the experiment is explained by our theoretical 
model \cite{Kullie:2015I}, the agreement with the FPI of Landsman 
\cite{Landsman:2014I,Landsman:2014II} treatment supports our theoretical result, 
especially because Landsman uses $\tau_0$ determined by the measurement to 
coarse-grain the probability distribution of the T-time. 

Indeed, B\"uttiker thought that ``reasonably we can only speak of a time duration 
if it is real and positive'' \cite{Muga:2008} (chap. 9). 
And Steinberg \cite{Muga:2008} (chap. 11) claimed that 
``the classical equivalence of a broad range of definitions cannot persist, 
for this one yield imaginary number, while most measurement techniques will be 
certain to yield positive values.'' To this end, one can suggest that our T-time 
is the real part of a (complex total) T-time in this experiment, the imaginary 
part would be the K-time or MT-time, i.e. the evolution time. However, the question 
is, how to understand such a situation in which the quantum dynamical evolution 
of the wave function has a different time scale from the real time of the tunneling 
process in the above mentioned sense of B\"uttiker and Steinberg. 
Because the time evolution in both procedure of Orlando and Zhao is done in  
a classical/laboratory frame, a possible answer is to consider the evolution 
in an intrinsic-time frame, e.g. as we sketched in subsec \ref{sssec:IEt}.

Furthermore, Sokolovski \cite{Sokolovski:1990}, 
\cite{Muga:2008} (chap. 7) claims in his FPI description that, no real time 
is associated with the tunneling, whereas the (real) T-time obtained by the 
coarse-grained FPI-probability distribution (with $\tau_0$ determined from 
the measurement) of Landsman \cite{Landsman:2014I,Landsman:2015} is in agreement 
with our real T-time.  Meaning, that the FPI treatment of Landsman avoids this 
circumstance of the intrinsic time frame, by the coarse-grain procedure based 
on the experiment (the measurement), which provides the internal-time values 
as discussed above in subsec \ref{sssec:IEt}. 
FPI treatment seems to benefit from both sides, the real and 
the imaginary, and possibly it can provide more insight in the above set 
question, and how to understand, in the tunneling process, the act of the real 
($\tau_{T,d},\tau_{T,sym}$) and the imaginary (e.g. $\tau_{K}$ of Zhao et al 
\cite{Zhao:2013}) parts. Similarly, it can also be fruitful the investigation 
of the time evolution of the wave function, but now under the consideration 
of the internal-time point of view (internal clock), this requires to choose 
a ‘reference’ point \cite{Maquet:2014}, which can be at best determined by the 
(natural) internal properties of the QM-system. 
One should mention at this point, that the time-of-arrival concept 
\cite{Allcock:1969I,Allcock:1969II,Allcock:1969III}, can hardly contribute 
to understanding this question (hopefully that is not completely excluded). 
Allcock concluded in his first paper \cite{Allcock:1969I}, that it is totally 
impossible to establish an ideal concept of arrival time for waveforms, 
which contain negative energy components.    
\vspace{-0.50cm}
\subsection{Conclusion}\label{ssec:con}
\vspace{-0.25cm}
We discussed in this work the T-time from different points of views, first the 
T-time in our treatment, which is in excellent agreement with the experiment 
\cite{Landsman:2014II,Eckle:2008,Eckle:2008s}, is a dynamical or intrinsic-type  
of time and represents a quantum clock, i.e. to observe the time form within 
the system. 
The MT-time as derived by \cite{Orlando:2014I,*Orlando:2014II} and the K-time are not capable 
of explaining the experimental result and represent external or laboratory type 
of time, when they are considered to be a real T-time.  
We gave a critique and remarks about the difference between the two views of 
the time, i.e. the intrinsic and the external type of time. 
Orlando et al. support their result with a numerical investigation, hence 
the puzzle of the T-time is still not completely understood. 
However, a similar K-time is obtained by Zhao et al \cite{Zhao:2013} from 
a numerical solution of the SEQ, but K-time is explained 
to be the imaginary part of a complex ionization time. 
MT-time can be brought under the same hat by taking a definition of Messiah 
\cite{Gray:2005,Messiah:1961}, as discussed in subsec \ref{sssec:what}. 
Hence we think that, an important point is rather how to observe the time in 
the system, which requires to choose a ‘reference’ point \cite{Maquet:2014}, 
which is determined by the internal properties of the QM-system.  
We provided an example form the special relativity (Muon decay), which shows 
a similar situation to the T-time and the controversial discussion (and views)
in attosecond and strong field experiments.   
In this respect we conclude that, the time in quantum mechanics can be, 
in more general fashion, classified in two types: intrinsic-type of time, 
dynamically connected to the system, and external/laboratory type of time, 
which is (dynamically) not connected to the system, despite that Briggs 
\cite{Briggs:2008,Maquet:2014} claims that there is no reason to introduce time for 
closed systems, he also claims that the time is classical, the $\Delta t$ 
enter in the TEUR must be associated with a classical measurement of the time 
(compare Hilgevoord in sec. \ref{ssec:int}). 
We think, that this view cannot persist, considering our today knowledge, 
view and insight of the T-time in attosecond experiments.  
A finial remark, it is likely that the transition form the 
intrinsic-type of time to the external-type of time of a quantum mechanical system 
(ionization, a transition from ''inside'' to ''outside'') is associated with 
a symmetry break of some property of the system. 
%
\end{document}